\documentclass{pasj00}
\begin{document}
\SetRunningHead{Author(s) in page-head}{Running Head}
%\Received{}%{yyyy/mm/dd}
%\Accepted{}%{yyyy/mm/dd}
%\Published{}%{yyyy/mm/dd}

\title{Detection of CO($J=1-0$) Emission from Barred Spiral Galaxies  at $z \sim 0.1$.}

%%% begin:list of authors
% Do NOT capitalize all letters in "textsc".
\author{
Kana \textsc{Matsui},\altaffilmark{1}
Kazuo \textsc{Sorai},\altaffilmark{1}
Yoshimasa {\sc Watanabe},\altaffilmark{2}
and
Nario {\sc Kuno}\altaffilmark{3,4}
}
\altaffiltext{1}{Department of Cosmosciences, Graduate School of Science, Hokkaido University, Kita-ku, Sapporo 060-0810}
\email{kmatsui@astro1.sci.hokudai.ac.jp, sorai@astro1.sci.hokudai.ac.jp}
\altaffiltext{2}{Department of Physics and Research Center for the Early Universe, The University of Tokyo, Bunkyo-ku, Tokyo 113-0033}
\email{nabe@taurus.phys.s.u-tokyo.ac.jp}
\and
\altaffiltext{3}{Nobeyama Radio Observatory, Minamimaki-mura, Minamisaku-gun, Nagano 384-1305}
\altaffiltext{4}{The Graduate University for Advanced Studies (SOKENDAI), 2-21-1 Osawa, Mitaka, Tokyo 181-0015}
\email{kuno@nro.nao.ac.jp}
%%% end:list of authors

%%% Please use the following style in case that sorting by 
%%% affilation is impossible. 
%
% \author{%
%   D-Firstname \textsc{D-Familyname}\altaffilmark{1}
%   E-Firstname \textsc{E-Familyname}\altaffilmark{1,2}
%   and
%   F-Firstname \textsc{F-Familyname}\altaffilmark{2}}
% \altaffiltext{1}{Address of Institute}
% \email{ddddd@xxx.xxx.xx.xx}
% \email{eeeee@xxx.xxx.xx.xx}
% \altaffiltext{2}{Address of Institute}

%% `\KeyWords{}' always has to be placed before `\maketitle'.
 %Do NOT move this preamble from here!

\maketitle

\begin{abstract}
We present the results of CO ($J=1-0$) observations towards nine barred spiral galaxies at $z=0.08-0.25$ using the 45-m telescope at Nobeyama Radio Observatory (NRO).
This survey is the first one specialized for barred spiral galaxies in this redshift range.
We detected CO emission from six out of nine galaxies, whose CO luminosity ($L_{\rm{CO}}'$) ranges $(1.09 - 10.8)\times 10^9\ \rm{K\ km\ s^{-1}\ pc^2}$.
These are the infrared (IR) dimmest galaxies that have ever been detected in CO at $z \sim 0.1$ to date.
They follow the $L_{\rm{CO}}'-L_{\rm{IR}}$ relation among local spiral galaxies, Luminous Infrared Galaxies (LIRGs), Ultra-Luminous Infrared Galaxies (ULIRGs) and Sub-millimeter Galaxies (SMGs).
Their $L_{\rm{CO}}'$ and $L_{\rm{IR}}$ are higher than that of local spiral galaxies which have been detected in CO so far, and $L_{\rm{IR}}/L'_{\rm{CO}}$, which is a measure of star formation efficiency, is comparable to or slightly higher than that of local ones.
This result suggests that these galaxies are forming stars more actively than local spirals galaxies simply because they have more fuel.
\end{abstract}

\section{Introduction}

Previous studies have revealed that almost 70 $\%$ of disk galaxies in the local universe have bars in their inner regions (e.g., \cite{esk00,elm04,jog04,men07,mar07,she08,agu09}).
This non-axisymmetric features is thought to influence the evolution of disk galaxies through redistributing angular momentum among the galaxy components, such as stars, gas and dark matter, and transferring gas to the central kiloparsec regions (e.g., Athanassoula 1992a, 1992b; \cite{sel93}).
Some observations have shown that molecular gas is more concentrated in central kiloparsec regions in barred spiral galaxies than non-barred ones \citep{sak99,she05,kun07}.

Recent mapping observations of CO ($J=1-0$) towards nearby spiral galaxies revealed diversity of molecular gas distributions, even though their morphologies are similar in optical images \citep{hel03,kun07}.
The molecular gas distributions are classified into four classes based on appearance in CO; molecular gas is
1) distributed all along the bar, 2) concentrated in the center and ends of the bar, 3) concentrated only in the center, and 4) little CO present in the bar \citep{wat10}.
These qualitative features of molecular gas distribution in bars can be quantified and the variety of distributions among galaxies are suggested to reflect the difference of the intrinsic nature and the evolutional stage of barred spiral galaxies \citep{wat10}.
It is very important to reveal the origin of the difference of molecular gas distributions, since molecular clouds are the very birth places of stars which shape galaxies.

In order to reveal whether there is a relation between the diversity of molecular gas distributions in barred spiral galaxies and their evolution, we need to investigate which type of distribution is the major trend in present and past epoch, and compare them each other.
However, it is impossible to make CO map of barred spiral galaxies at $z$\hspace{0.3em}\raisebox{0.4ex}{$>$}\hspace{-0.75em}\raisebox{-.7ex}{$\sim$}\hspace{0.3em}0.1 in comparable resolution with that for nearby galaxies,\hspace{0.3em}\raisebox{0.4ex}{$<$}\hspace{-0.75em}\raisebox{-.7ex}{$\sim$}\hspace{0.3em}1 kpc, using the existing instruments.
Observations with the Atacama Large Millimeter/submillimeter Array (ALMA), which is expected to have higher resolution and sensitivities, will shed light on the appearance of CO in galaxies at $z$\hspace{0.3em}\raisebox{0.4ex}{$>$}\hspace{-0.75em}\raisebox{-.7ex}{$\sim$}\hspace{0.3em}0.1.

Although a number of CO observations toward galaxies at $z$\hspace{0.3em}\raisebox{0.4ex}{$>$}\hspace{-0.75em}\raisebox{-.7ex}{$\sim$}\hspace{0.3em}0.1 have been made so far, they are limited to Luminous Infrared Galaxies (LIRGs), Ultra-Luminous Infrared Galaxies (ULIRGs) and Sub-millimeter Galaxies (SMGs).
Moreover, these studies are not concerned with the morphology of galaxies (Scoville et al. 1993, 2003; Solomon et al. 1997, 2005; \cite{lo99,tut00}; Evans et al. 1999, 2002, 2005, 2006, 2009; \cite{gao04,dad10,gea11,com11}).
Meanwhile, studies concerning the morphology of galaxies have so far been limited to nearby galaxies.
It is now possible to see the morphology, especially bars, of galaxies in optical or near infrared (NIR), and also to detect CO emission at 0.1\hspace{0.3em}\raisebox{0.4ex}{$<$}\hspace{-0.75em}\raisebox{-.7ex}{$\sim$}\hspace{0.3em}$z$\hspace{0.3em}\raisebox{0.4ex}{$<$}\hspace{-0.75em}\raisebox{-.7ex}{$\sim$}\hspace{0.3em}1.0, thanks to the progression of observational equipment, even though the resolution obtained in a reasonable observing time can not match that of local observations.
This allows us to perform for the first time the systematic CO observations towards barred spiral galaxies in this redshift range.
Hence we performed CO ($J=1-0$) pointing observations towards barred spiral galaxies at $z$\hspace{0.3em}\raisebox{0.4ex}{$>$}\hspace{-0.75em}\raisebox{-.7ex}{$\sim$}\hspace{0.3em}0.1 using the 45-m telescope at Nobeyama Radio Observatory (NRO)\footnote{The 45-m radio telescope is operated by Nobeyama Radio Observatory, a branch of National Astronomical Observatory of Japan.}.
The results from this survey will provide us the molecular gas mass of normal galaxies in this redshift range for the first time and also act as a sample selector for upcoming ALMA observation.
As a first step, we observed nine barred spiral galaxies at $z \sim$ 0.1.

We assume the standard $\Lambda-$CDM cosmology with $H_0 = 71\ \rm{km\ s^{-1}\ Mpc^{-1}}$, $\Omega_{M}=0.27$, $\Omega_{\Lambda}=0.73$ \citep{koma09} through this paper.

%%%%%%%%%%%%%%%%%%%%%%%%%%%%%%%%%%%%%%%%%%%%%%%%
%%%%%%%%%%%%%%%%%%%%%%%%%%%%%%%%%%%%%%%%%%%%%%%%
\section{Sample Selection}

Our purpose of these observations is to search for barred spiral galaxies around $z \sim 0.1$ which will be bright enough in CO to see their molecular gas distributions in bars through upcoming ALMA observation.
This purpose requires our sample to satisfy following criteria:
\begin{enumerate}
\item barred spiral galaxies,
\item redshift of $z\sim0.1$,
\item detectable CO ($J=1-0$) emission.
\end{enumerate}

We need high resolution optical or NIR images to determine whether galaxies have a bar or not for the first criterion.
Thus, the samples are selected from Cosmic Evolution Survey 2 $\rm deg^2$ field (COSMOS: \cite{sco07}).
We picked up sample galaxies in two ways; through Structured Query Language (SQL) query in SDSS/SkyServer\footnote{SDSS/SkyServer: http://cas.sdss.org/astro/en/} and  from ZCOSMOS \citep{lil09} catalog\footnote{ZCOSMOS DR2: http://archive.eso.org/cms/eso-data/data-packages/zcosmos-data-release-dr2/}.

For selecting galaxies through SkyServer, we specified the sample selection area which is $2\rm \ deg^2$ square centered on $(\alpha, \delta)_{J2000}$ = $(10^{\rm{h}}\ 00^{\rm{m}}\ 28^{\rm{s}}.6, +02^\circ \ 12'\ 21''.0)$, which is the center of the COSMOS field, and the redshift range 0.08 $< z <$ 0.25.
In order to pick up disk galaxies and to exclude galaxies too inclined to speculate on their face-on morphologies, we imposed following criteria:
\begin{enumerate}
\item the likelihood of the exponential fit is greater than the likelihood of  the de Vaucouleours profile fit,
\item the inclination angle obtained from the axis ratio of exponential fit is less than $60^\circ$.
\end{enumerate}
We used $r$ band data for sample selection, since it achieved the best sensitivity (22.2 mag) among all five SDSS bands.
This left us with 46 galaxies that satisfied these criteria.

In the case of selection from ZCOSMOS catalog, we picked up galaxies by imposing only redshift criterion, since  ZCOSMOS catalog does not contain the information on the morphology of galaxies.
1,064 galaxies remained that still include high inclined disk, elliptical and irregular galaxies. 

\begin{figure}
   \begin{center}
      \FigureFile(85mm,85mm){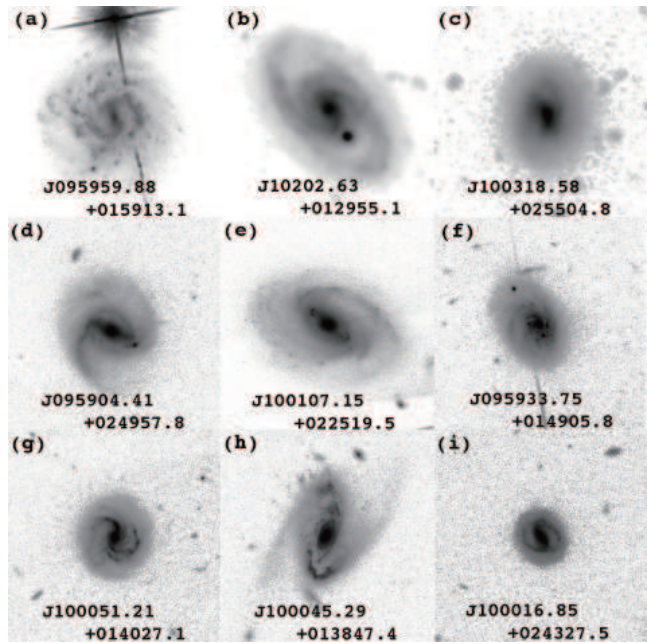}
   \end{center}
   \caption{Optical $I$ band ((a), (d), (e), (f), (g), (h) and (i) taken with HST/ACS: 
   \cite{koek07}
   ) or $B$ band ((b) and (c) taken with Subaru) images of sample galaxies. Image size is 18 arcsec $\times$ 18 arcsec.}
   \label{fig-1}
\end{figure}

\begin{table*}
\begin{center}
\begin{tabular}{cccccc}
\hline\hline
SDSS Name & $\alpha_{J2000}$& $\delta_{J2000}$ & redshift & $ L_{\rm IR}$&Central \\
& $\rm ^h\ \ ^m\ \ ^s$& $^\circ\ \ \ '\ \ \ ''$ & & {\footnotesize@($10^{10}\ L_\odot$})&activity\footnotemark[$\ast$]\\
\hline
J095959.88+015913.1 & 09 59 59.9 & +01 59 13.0 & 0.080346 & $2.13\pm0.15$\footnotemark[$\ast\ast$]&HII\\
J100202.63+012955.1 & 10 02 02.7 & +01 29 57.5 & 0.098167 & $3.08\pm0.45$ &AGN\\
J100318.58+025504.8 & 10 03 18.6 & +02 55 04.8 & 0.104808 & $3.58\pm0.56$ &HII\\
J095904.41+024957.8 & 09 59 04.4 & +02 49 57.9 & 0.119338 & $4.44\pm0.61$&HII\\
J100107.15+022519.5 & 10 01 07.2 & +02 25 19.6 & 0.1215 & $4.18\pm0.68$ &...\\
J095933.75+014905.8 & 09 59 33.7 & +01 49 05.9 & 0.132985 & $12.4\pm1.4$&HII\\
J100051.21+014027.1 & 10 00 51.2 & +01 40 27.2& 0.166130 & $14.5\pm1.6$&HII\\
J100045.29+013847.4 & 10 00 45.2 & +01 38 47.5& 0.220522 & $18.5\pm3.0$&transition\\
J100016.85+024327.5 & 10 00 16.9 & +02 43 27.5& 0.2500 & $22.3\pm2.8$&...\\
\hline
\end{tabular}
\caption{Sample galaxies.\label{tbl-1} \footnotemark[$\ast$]Classified using BPT diagram \citep{bal81}, and criteria defined in \citet{kew06}. \footnotemark[$\ast\ast$]Estimated from only $S_{24}$, $S_{70}$ and assumed $\nu_{70} L_{70}=\nu_{160} L_{160}$, since this galaxy does not have $S_{160}$ data.}
\end{center}
\end{table*}

Next, we cross-matched both sets of galaxy candidates with S-COSMOS \citep{san07} data catalog through NASA/IPAC Infrared Science Archive (IRSA).\footnote{IRSA General Catalog Query Engine: http://irsa.ipac.caltech.edu/applications/Gator/}
We picked up IR bright galaxies with 24$\mu$m flux $S_{24} > 2.0\ \rm{mJy}$, expected to satisfy the third criterion from the $L_{\rm{IR}}-L'_{\rm{CO}}$ relation among various galaxies reported in the wide redshift range so far (\cite{you84,san85,tut00,yao03,gre05}). 

IR luminosity, $L_{\rm IR}$ was estimated using Spitzer/MIPS data as
\begin{equation}
L_{\rm{IR}} = 1.559\  \nu_{24} L_{24} + 0.7686\ \nu_{70} L_{70} + 1.347\ \nu_{160} L_{160}
\label{LIR}
\end{equation}
where $L_{24}$, $L_{70}$ and $L_{160}$ are luminosities at 24$\mu$m, 70$\mu$m and 160$\mu$m, respectively \citep{dal02}.

Finally, we retrieved optical or NIR images taken with HST, Subaru Telescope or Canada France Hawaii Telescope as a part of COSMOS project through COSMOS Cutouts\footnote{COSMOS Cutouts: $\rm http://irsa.ipac.caltech.edu/data/COSMOS/index\_cutouts.html$} and checked whether each galaxy had a bar or not (the first criterion).
This resulted in nine galaxies remained.
Five out of nine sample galaxies are classified as normal and four as LIRGs in terms of IR luminosity.
Normal-type samples are the ones with the IR dimmest galaxies that are detected in CO at this redshift.
Optical images of these galaxies are shown in figure \ref{fig-1} and parameters of the sample galaxies are listed in table \ref{tbl-1}.

%%%%%%%%%%%%%%%%%%%%%%%%%%%%%%%%%%%%%%%%%%%%%%%%
%%%%%%%%%%%%%%%%%%%%%%%%%%%%%%%%%%%%%%%%%%%%%%%%
\section{Observation}

We observed nine barred spiral galaxies at $z \sim 0.1$ in the CO ($J=1-0$) line from January to May in 2010 and 2011 using the 45-m telescope in NRO.
The line frequency was shifted to between 92.217 GHz and 106.699 GHz from the rest frequency of 115.271 GHz according to the redshifts of the sample galaxies.
We used waveguide-type dual-polarization sideband-separating SIS receivers, T100H/V \citep{nak08} and 1024 channel digital autocorrelators with a frequency coverage of 512 MHz as a backend \citep{sor00}.
The image rejection ratios (IRRs) of T100H/V were measured at each observing frequency per day and were $9 - 20$ dB through the observations.
The system noise temperature, $T_{\rm{sys}}$, was typically $150 - 240$ K during our observations.
Telescope pointing was checked every 50 minutes through observing SiO maser source, R leo, at 43 GHz.
We used the data which was observed with pointing accuracy higher than 5 arcsec.
The main beam size, $\theta_{\rm{B}}$ is $\sim$ 18 arcsec around 100 GHz, which corresponds to $\sim$ 27 kpc at the smallest redshift of sample galaxies ($z=0.08$), and to $\sim$ 70 kpc at the largest ($z=0.25$).
All data were calibrated by the standard chopper wheel method, and converted from antenna temperature $T^*_a$ scale into main-beam brightness temperature $T_{\rm{mb}}=T^*_a/\eta_{\rm{mb}}$.
Here the main beam efficiency $\eta_{\rm{mb}}$ was 0.4\footnote{Conversion factor from K to Jy around $92-107$ GHz, which is the range of observing frequencies, is estimated to be $\sim$ 5.8 Jy/K.}.

Data reduction was done using the NEWSTAR software, which was developed by the NRO based on the Astronomical Image Processing System (AIPS) package. 
All data were scaled by the following factor $f$ \citep{ker01},
\begin{equation}
f = 1 + \frac{1}{10^{\rm IRR/10}}.
\end{equation}

We flagged all data with r.m.s. $>$ 0.075 K in $T_{\rm{mb}}$ scale at 200 $\rm km\ s^{-1}$ resolution to exclude bad baseline spectra, and then baselines were subtracted by linear fitting.
Integration time of each galaxy was $2 - 8$ hours summed up for the both polarizations, and the typical r.m.s. noise temperature was in the range of $1.0 - 2.5$ mK in $T_{\rm{mb}}$ scale after binning up to $40-50$ $\rm km\ s^{-1}$ resolution.
Integrated intensity, $I_{\rm{CO}}$, was calculated according to $I_{\rm{CO}} = \int T_{\rm{mb}}\ dv$.
The error of $I_{\rm{CO}}$, $\Delta I_{\rm{CO}}$, was estimated as \citep{sag93}
\begin{equation}
\Delta I_{\rm{CO}} = \sqrt{\Delta V^2_{\rm e} \left( \frac{\Delta v}{\Delta V_{\rm e}} \right) (T_{{\rm r.m.s.}})^2 + \Delta V^2_{\rm e} \left( \frac{\Delta v}{\Delta V_{\rm b}} \right) (T_{{\rm r.m.s.}})^2},
\end{equation}
where $\Delta V_{\rm e}$ is the full line width within which the integrated intensity was calculated, and $\Delta V_{\rm b}$ is the velocity range over which the baseline was fitted.
$T_{{\rm r.m.s.}}$ is the r.m.s. noise temperatures at velocity resolution of $\Delta v$, which is the velocity resolution of final spectrum.
The first term in the square root represents the error from r.m.s. noise and the second term represents that from the error of the baseline determination.
In the case where the spectra has a signal-to-noise ratio (S/N) less than 3, the upper limit of 3 $\Delta I_{\rm{CO}}$ was adopted as $I_{\rm{CO}}$.

CO line luminosity was calculated from the integrated intensity and beam solid angle $\Omega_{b}=\pi \theta_{\rm{B}}^2/4 \ln{2}$ ($\theta_{\rm{B}}$ in radians):
\begin{equation}
L_{\rm{CO}}' = \frac{\Omega_{b} I_{\rm{CO}} D_L^2}{(1+z)^3}\ \ (\rm{K\ km\ s^{-1}\ pc^2}),
\end{equation}
where $D_L$ is a luminosity distance formulated by cosmological parameters such as Hubble constant $H_0$, matter density parameter $\Omega_M$, cosmological constant density parameter $\Omega_\Lambda$ as:
\begin{eqnarray}
D_L = (1+z)\frac{c}{H_0} \int^z_0 \frac{dz} {\left[ \Omega_{M} (1+z)^3 + \Omega_\Lambda \right]^{1/2}}.
\end{eqnarray}

%%%%%%%%%%%%%%%%%%%%%%%%%%%%%%%%%%%%%%%%%%%%%%%%
%%%%%%%%%%%%%%%%%%%%%%%%%%%%%%%%%%%%%%%%%%%%%%%%
\section{Results and Discussions}

\begin{figure*}
   \begin{center}
      \FigureFile(160mm,140mm){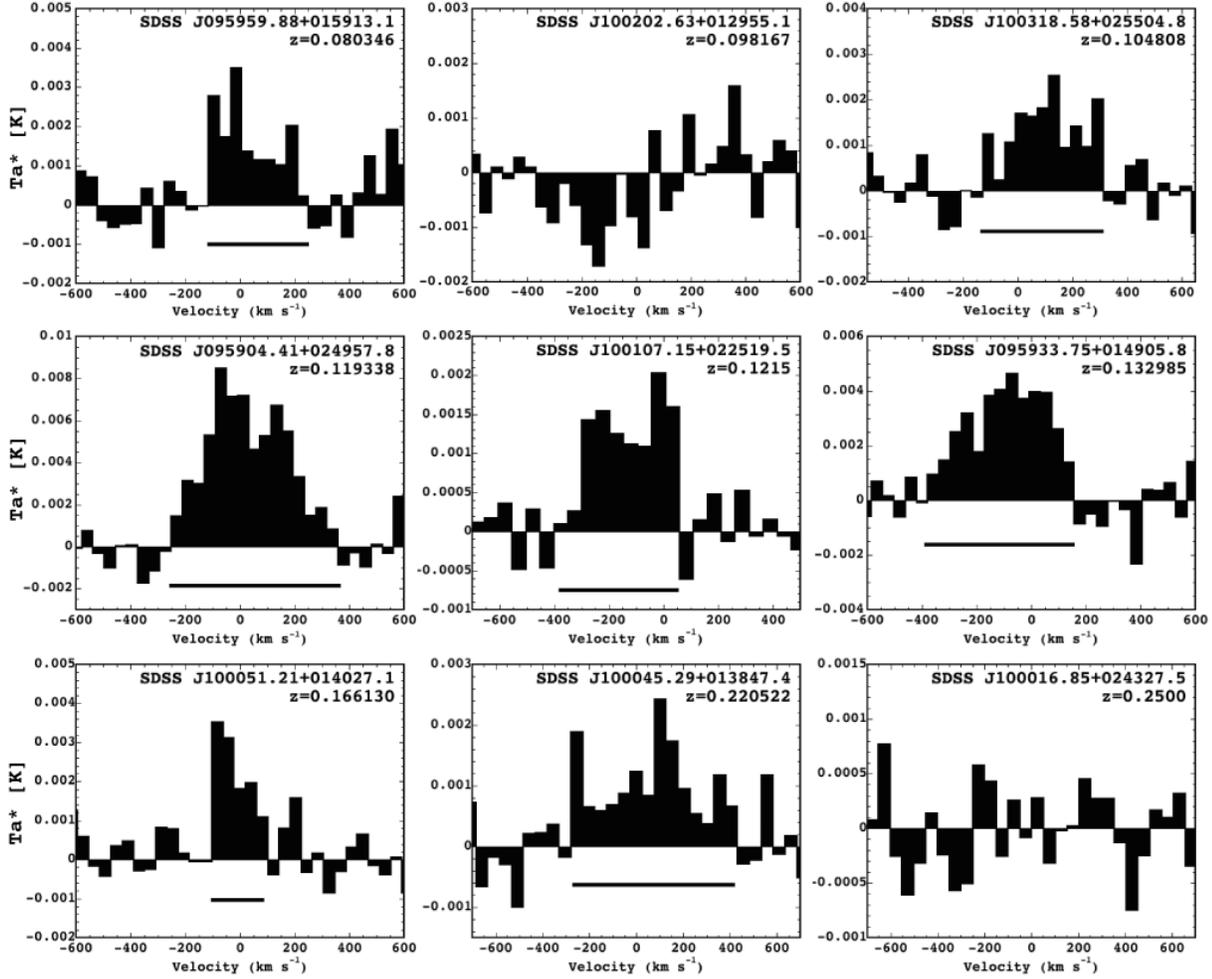}
   \end{center}
   \caption{CO spectra taken with the 45-m telescope. All spectra are binned into $40-50$ $\rm km\ s^{-1}$ resolution. Name and redshift of each galaxy are shown in the upper right corner of the spectrum. Heavy line in each spectrum, except for SDSS J100202.63+012955.1 and SDSS J100016.85+024327.5 (S/N $<$ 3), indicates the full width at zero intensity of emission line.}
    \label{fig-2}
\end{figure*}

The obtained spectra are shown in figure \ref{fig-2}.
We detected CO emission from six out of nine galaxies (S/N $>$ 5 in $I_{\rm{CO}}$), marginally from SDSS J100045.29+013847.4 (S/N$\sim$4.4), while the remaining  SDSS J100202.63+012955.1 and SDSS J100016.85+024327.5 did not show CO emission clearly (S/N $<$ 3).
CO line luminosity of our sample galaxies ranges from $1.09 \times 10^9$ (SDSS J095959.88+015913.1) to $1.08 \times 10^{10}$ (SDSS J100051.21+014027.1) $\rm{K\ km\ s^{-1}\ pc^2}$, and average value is $\sim 5 \times10^9$ $\rm{K\ km\ s^{-1}\ pc^2}$, which is several times higher than that of local spiral galaxies.
If we applied the local CO-to-$\rm H_2$ conversion factor of $X_{\rm{CO-H_2}}=1.0\times 10^{20}\ \rm{cm^{-2}\ (K\ km\ s^{-1})^{-1}}$ \citep{nak95} to our sample galaxies, molecular gas mass would be in the range of $1.75\times 10^9 - 1.74\times 10^{10}\ M_\odot$.
CO integrated intensity, full width at half maximum (FWHM) of spectrum, CO line luminosity, molecular gas mass, and $L_{\rm{IR}}/L_{\rm{CO}}'$ of our sample galaxies are summarized in table \ref{tbl-2}.

\begin{table*}
\begin{center}
\caption{Summary of observations and derived parameters.\label{tbl-2}}
\begin{tabular}{cccccc}
\hline\hline
SDSS Name & $I_{\rm{CO}}$ & FWHM & $L_{\rm{CO}}'$ & $M_{\rm{H2}}$ & $L_{\rm{IR}}/L_{\rm{CO}}'$\\
 & {\footnotesize $\rm{K\ km\ s^{-1}}$} & {\footnotesize $\rm{km\ s^{-1}}$} & {\footnotesize $10^9\ \rm{K\ km\ s^{-1}\ pc^2}$} & {\footnotesize $10^9\ M_\odot$} & {\footnotesize $L_\odot\ (\rm{K\ km\ s^{-1}\ pc^2})^{-1}$}\\
\hline
J095959.88+015913.1 & $1.46\pm0.26$ & 243 & $1.09\pm0.19$ & $1.75\pm0.31$ & $19.5\pm3.7$\\
J100202.63+012955.1 & $< 0.67$ & ... & $< 0.75$ & $< 1.20$ & $> 41.1$\\
J100318.58+025504.8 & $1.55\pm0.21$ & 303 & $1.97\pm0.25$ & $3.16\pm0.43$ & $18.1\pm3.8$\\
J095904.41+024957.8 & $6.56\pm0.52$ & 341 & $10.8\pm0.87$ & $17.4\pm1.4$ & $4.10\pm0.65$\\
J100107.15+022519.5 & $1.25\pm0.19$ & 322 & $2.15\pm0.32$ & $3.44\pm0.52$ & $19.5\pm4.3$\\
J095933.75+014905.8 & $3.84\pm0.38$ & 329 & $7.88\pm0.79$ & $12.6\pm1.3$ & $15.8\pm2.4$\\
J100051.21+014027.1 & $1.14\pm0.12$ & 141 & $3.64\pm0.39$ & $5.84\pm0.62$ & $39.7\pm6.2$\\
J100045.29+013847.4 & $1.82\pm0.41$ & 533 & $10.2\pm2.30$ & $16.3\pm3.7$ & $18.2\pm5.1$\\
J100016.85+024327.5 & $<0.42$ & ... & $<3.01$ & $<4.83$ & $>74.0$\\
\hline
\end{tabular}
\end{center}
\end{table*}

\subsection{Validity of estimation of galactic global value of $L_{\rm{CO}}'$ and $L_{\rm{IR}}$}
We estimated $L_{\rm{IR}}$ from 24, 70 and 160 $\mu$m fluxes listed in the S-COSMOS catalog.
COSMOS is the survey for distant sources which are expected to be point sources with respect to the size of Point Spread Function (PSF).
For the three bands of MIPS, the PSF is 6 arcsec for 24 $\mu$m, 18 arcsec for 70$\mu$m, and 40 arcsec for 160 $\mu$m.
The largest Petrosian radius in $r$-band, $R_{{\rm Pet}, r}$ among our sample galaxies is $5.45\pm0.24$ arcsec (SDSS J100045.29+013847.4), and it has been shown that $2R_{{\rm Pet}, r}$ encompasses almost all of the total light from an exponential disk galaxy \citep{str02}.
The optical size of this largest sample is estimated to be $\sim 22$ arcsec ($>6$ arcsec and $>18$ arcsec).
Therefore there is a possibility that the fluxes calculated as point sources may be underestimated, even though it is unclear whether the galaxy extent of our samples in MIPS bands is comparable to the optical size or not.
We examined this by comparing the $L_{\rm{IR}}$ estimated from the catalog fluxes with the one estimated from images.
For the $L_{\rm{IR}}$ estimation from MIPS images of galaxies, we calculated the total fluxes included in the circles with diameter of $4-5$ times FWHM of PSFs of each band.
As a result, we got an almost linear relation between two $L_{\rm{IR}}$s estimated in two different ways.
Thus it can be concluded that the degree to which the galactic global $L_{\rm{IR}}$ calculated from cataloged fluxes is underestimated is negligibly small.

In case of $L_{\rm{CO}}'$, we estimated the galactic global value from this observation using a beam size of the telescope that was slightly larger than or comparable to the optical size of observed object.
In order to estimate the degree to which we underestimate $L_{\rm{CO}}'$ with respect to the relative size of the CO extent in galaxies, compared to the telescope beam size, we assumed a simple model.
We assumed an exponential distribution of CO with several scale lengths and the gaussian telescope beam profile, and calculated the ratio of observed intensity to the true value as a function of the relative size of telescope beam to the CO extent.

\begin{figure}
   \begin{center}
      \FigureFile(80mm,80mm){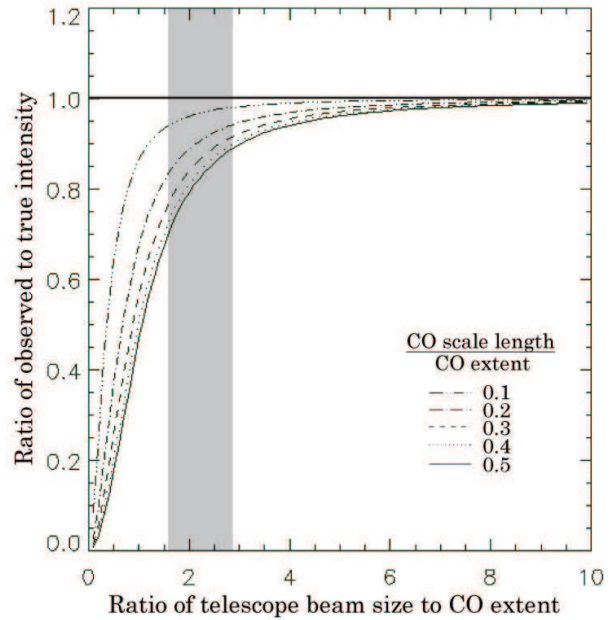}
   \end{center}
   \caption{The degree of recovery of true intensity as a function of the relative size of telescope beam to CO extent. Five lines in this plot represent the results with different ratio of CO scale length to the extent of CO, where 0.1, 0.2, 0.3, 0.4, and 0.5 from top down. Shaded area is the expected range of horizontal axis of CO-detected galaxies in this observation, $\sim 1.6 - 2.8$.}
   \label{fig-3}
\end{figure}

The result is shown in figure \ref{fig-3}.
Five lines in this plot represent the results with different ratio of CO scale length to the extent of CO with 0.1, 0.2, 0.3, 0.4, and 0.5 from top down.
\citet{you95} has shown that the CO extent is almost half of the optical size of galaxies, and its scale length is about one-fifth of the optical radius.
It has been also shown that the degree of central concentration of CO of barred spiral galaxies is higher than that of non-barred ones as mentioned above \citep{sak99}. 
Therefore this ratio is estimated to be \hspace{0.3em}\raisebox{0.4ex}{$<$}\hspace{-0.75em}\raisebox{-.7ex}{$\sim$}\hspace{0.3em}0.4.
In this plot, we can see that the larger the relative size of telescope beam is to the CO extent, the more precisely the galactic global intensity is recovered.
Moreover, the degree of recovery depends on the scale length of CO distribution, i.e., the observed intensity is close to the true value with the central concentration of the CO.

Since the largest optical size among sample galaxies is about $\sim 22$ arcsec, the gas extent is estimated to be $\sim 11$ arcsec.
In that case, the ratio of the size of telescope beam ($\sim 18$ arcsec) to the CO extent is $\sim 1.6$, thus the degree of intensity loss is estimated to be \hspace{0.3em}\raisebox{0.4ex}{$<$}\hspace{-0.75em}\raisebox{-.7ex}{$\sim$}\hspace{0.3em}27\%.
Meanwhile the error of the intensity in this galaxy is $\sim23$\%, which is comparable to the loss caused by the relative size of CO extent to the size of telescope beam.
Therefore, although the galactic global $L_{\rm{CO}}'$ obtained by this observation may be underestimated in some degree, the intensity loss is expected to be relatively small.

\subsection{Global $L_{\rm{CO}}'-L_{\rm{IR}}$ relation and $L_{\rm{IR}}/L_{\rm{CO}}'$ as a function of redshift}
It is known that there is correlation between CO line luminosity $L_{\rm{CO}}'$ and IR luminosity $L_{\rm{IR}}$, which has been reported over wide redshift ranges (e.g., \cite{you84,san85,tut00,yao03,gre05}).
This relation means that molecular gas mass ($M_{\rm H_2}$) correlates with star formation rate (SFR).
However, CO observations at $z$\hspace{0.3em}\raisebox{0.4ex}{$>$}\hspace{-0.75em}\raisebox{-.7ex}{$\sim$}\hspace{0.3em}0.1 are currently limited to LIRGs, ULIRGs and SMGs, which are the extremely IR (in the rest frame) luminous objects in the universe.
It still remains unclear whether this relation holds in normal galaxies ($L_{\rm{IR}} < 10^{11}\ L_\odot$) at $z\sim0.1$.

\begin{figure*}
   \begin{center}
      \FigureFile(160mm,85mm){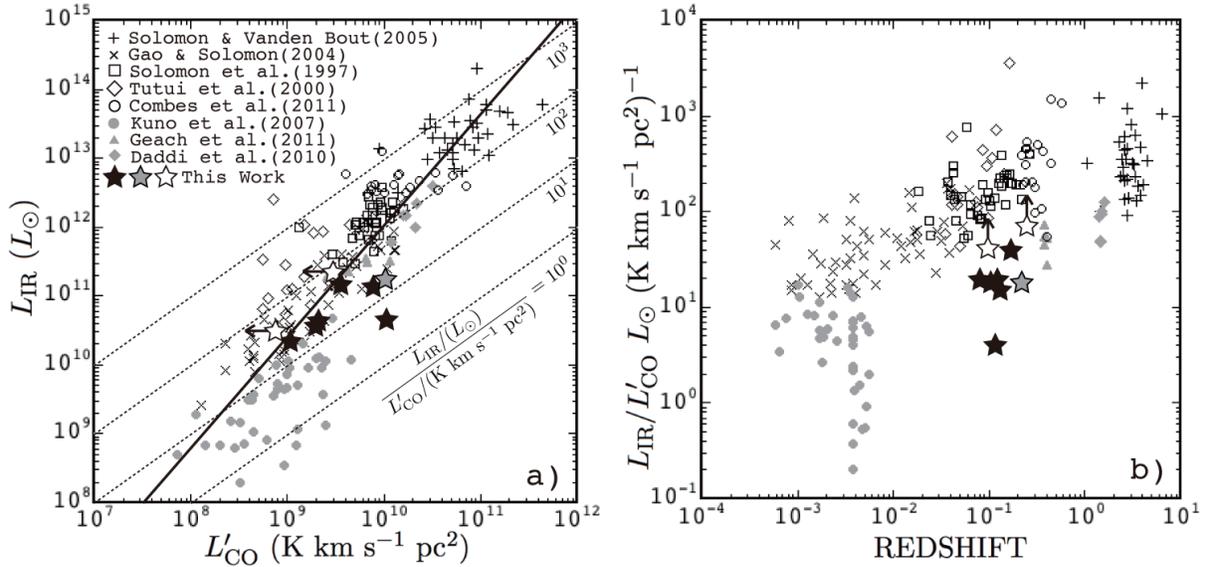}
   \end{center}
   \caption{(a) $L_{\rm{CO}}'$ vs $L_{\rm{IR}}$ of our sample galaxies as well as the data from previous studies. Cross: SMGs, x-mark: local normal galaxies and LIRGs, open-box: LIRGs and ULIRGs, open-diamond: normal galaxies, LIRGs and ULIRGs, grey-circle: local spiral galaxies, grey-triangle: disk galaxies at $z\sim0.4$, and grey-diamond: BzK galaxies. Black star: the CO-detected galaxies (S/N$>$5), gray star: the marginally CO-detected galaxies (S/N$\sim$4.4), and white star: not CO-detected galaxies (S/N$<$3) in this work. The errors of our data are as large as the size of star symbols. Four parallel dotted lines represent different $L_{\rm{IR}}/L_{\rm{CO}}'$ ratio of $10^3$, $10^2$, $10^1$ and $10^0$ $L_{\odot}\ (\rm{K\ km\ s^{-1}\ pc^2})^{-1}$. The solid line represents the least-square fit to all data, where $\log{L_{\rm{IR}}}=(1.61\pm0.06)\log{L_{\rm{CO}}'}-(4.11\pm0.54)$.  (b) Redshift vs $L_{\rm{IR}}/L_{\rm{CO}}'$. Symbols are the same as (a).}
   \label{fig-4}
\end{figure*}

We plot $L_{\rm{CO}}'$ versus $L_{\rm{IR}}$ for our sample galaxies as well as the data from previous studies in figure \ref{fig-4} (a).
Here, we used $L_{\rm{CO}}'$ and $L_{\rm{IR}}$ rather than $M_{\rm{H_{2}}}$ and SFR, since the former values are obtained through observations without any assumptions like conversion factors, which may be different between normal and extremely luminous galaxies.
We used published luminosities of $z<1$ objects; LIRGs and ULIRGs from \citet{sol97} and \citet{tut00}, normal galaxies and LIRGs from \citet{gao04}, and disk galaxies at $z\sim0.4$ from \citet{gea11}, and those of $z>1$ objects; SMGs from \citet{sol05}, and $BzK$s from \citet{dad10}.
For local spiral galaxies, we used the cube data of Nobeyama CO Atlas \citep{kun07}, summing up spectra of each galaxy as if they were observed through a telescope beam covering the entire galaxy as in this observation, and estimated their global $L_{\rm{CO}}'$.
Their global $L_{\rm{IR}}$ were estimated using IRAS data as
\begin{equation}
L_{\rm{IR}}=3.65\times10^5\ (2.58\ S_{60}+S_{100})\ D^2\ \ (L_\odot)
\label{Lir_IRAS}
\end{equation}
where $S_{60}$ and $S_{100}$ are the fluxes at 60$\mu$m and 100$\mu$m in Jy and $D$ is the distance in Mpc \citep{dev90}.
Hereafter, we use the term $"$normal spiral galaxies$"$ for local spirals which have been detected in CO even though there may be more IR-luminous galaxies in the local universe.

Distant luminous galaxies are often observed through only higher transition lines of CO, thus we need to estimate $L'_{\rm{CO} (\it{J}=\rm{1-0})}$ from their luminosities of  $J>2$.
There are some studies showing that $r_{21}=L'_{\rm{CO} (\it{J}=\rm{2-1})}/L'_{\rm{CO} (\it{J}=\rm{1-0})}$, and similarly $r_{32}$ and $r_{43}$ are less than unity ($r_{21}=0.9$ for nearby normal and IR-bright galaxies: \cite{bra92,aal95}, $r_{32}=0.64$ for nearby starbursts: \cite{dev94}, $r_{32}=0.50$ for $BzK$s: \cite{dan09}, $r_{43}=0.45$ for radio galaxies at $z\sim3-4$: \cite{pap00}).
Since they are consistently within this range and also recent CO($J=1-0$) observations towards quasar host galaxies at $z>2$ have shown that these distant-luminous galaxies had $r_{32}=1$ \citep{rie11}, for simplicity we assumed $r_{21}=r_{32}=r_{43}=1$, which are the thermalized optically thick CO emissions, and estimated $L'_{\rm{CO} (\it{J}=\rm{1-0})}$.
Four parallel dotted lines in figure \ref{fig-4} (a) represent different $L_{\rm{IR}}/L_{\rm{CO}}'$ ratio of $10^3$, $10^2$, $10^1$ and $10^0$ $L_{\odot}\ (\rm{K\ km\ s^{-1}\ pc^2})^{-1}$.

Our sample galaxies seem to follow the $L_{\rm{CO}}' - L_{\rm{IR}}$ relation throughout the data set, including local spiral galaxies, LIRGs, ULIRGs and SMGs even though there is a certain degree of scatters in figure \ref{fig-4} (a).
In this plot, our CO-detected samples seem to be distributed between local spiral galaxies and LIRGs, which implies that they have a larger amount of molecular gas and higher SFR than local spiral galaxies.
In figure \ref{fig-4} (b), we plot the redshift versus $L_{\rm{IR}}/L_{\rm{CO}}'$, which is a measure of star formation efficiency (SFE).
$L_{\rm{IR}}/L_{\rm{CO}}'$ of our CO-detected sample are in the range of $4.1 - 40$ $L_\odot\ (\rm{K\ km\ s^{-1}\ pc^{2}})^{-1}$ which are lower than other galaxies that have ever been detected in CO at $z\sim0.1$.

In figure \ref{fig-4} (a), focusing on a fixed $L_{\rm{CO}}'$ especially around $10^9\ \rm{K\ km\ s^{-1}\ pc^2}$, there is a scatter of nearly two-orders of magnitude of $L_{\rm{IR}}$.
This scatter reflects the difference of SFE, which is indicated as different lines of constant $L_{\rm{IR}}/L_{\rm{CO}}'$, of each galaxy in this plot.
Normal galaxies ($L_{\rm{IR}} < 10^{11}\ L_\odot$) are distributed in $L_{\rm{IR}}/L_{\rm{CO}}' < 10^2\ L_\odot\ (\rm{K\ km\ s^{-1}\ pc^2})^{-1}$ as well as our sample galaxies, while extremely IR-luminous galaxies ($L_{\rm{IR}} > 10^{12}\ L_\odot$) are distributed in $L_{\rm{IR}}/L_{\rm{CO}}' > 10^2\ L_\odot\ (\rm{K\ km\ s^{-1}\ pc^2})^{-1}$.
This means that some galaxies experience a normal star formation as occurring in the disk region of local galaxies while others experience a burst-like one, even with the same amount of molecular gas.

Combined with these results, our sample galaxies in $z\sim0.1$ seems to have a higher star formation activity than the normal spiral galaxies, not because their SFE is quite different but because they have more molecular gas as fuel for the star formation.

\subsection{Origin of scatter in $L_{\rm{IR}}/L_{\rm{CO}}'$}
In figure \ref{fig-4} (b), our sample galaxies seem to be distributed in a relatively wide range of $L_{\rm{IR}}/L_{\rm{CO}}'$, taking into account non-detected samples.
It is expected that there are normal galaxies at higher redshift, but CO observation of higher redshift has been limited so far towards LIRGs or ULIRGs as mentioned.
Our sample galaxies with lower value of $L_{\rm{IR}}/L_{\rm{CO}}'$ are normal-type galaxies as shown above.
On the other hand, our sample also contains galaxies with relatively high $L_{\rm{IR}}/L_{\rm{CO}}'$.
We classified sample galaxies picked up from SDSS in terms of activity of the central region of galaxies (HII, transition or AGN) using flux ratio of optical emission lines listed in the SDSS catalog.
As a result, SDSS J100202.63+012955.1 is classified as an AGN, SDSS J100045.29+013847.4 is in the transition phase, and the others are classified as HII nucleus (see Table \ref{tbl-1}).
This indicate that the contribution from the AGN to their $L_{\rm{IR}}$ may not be negligible.
For galaxies whose $L_{\rm{IR}}$ is expected to show mainly their star formation activity, active starbursts which typically occur in the central region of the galaxy may affect the galactic global value of $L_{\rm{IR}}/L_{\rm{CO}}'$.
Hence it can be expected that $L_{\rm{IR}}/L_{\rm{CO}}'$ value in the center of starburst galaxies contributes more to the galactic global value than that of normal galaxies.
We cannot separate the central region from the disk in the IR or CO image of our sample galaxies at $z\sim0.1$ due to low resolution of these data, especially the beam size of CO observation was larger than the spatial extent of sample galaxies.
Thus we examined the contribution from central $L_{\rm{IR}}/L_{\rm{CO}}'$ to the galactic global one using the data of local spiral galaxies listed in Nobeyama CO Atlas (\cite{kun07}, hereafter Nobeyama samples).
The CO map is from the atlas, and the IR map from Spitzer/MIPS 24 $\mu$m.
Although $L_{\rm{IR}}$ of Nobeyama samples in figure \ref{fig-4} were derived using IRAS data and equation (\ref{Lir_IRAS}), we used MIPS 24$\mu$m map in this sub-section since the spatial resolution of IRAS (60, 100 $\mu$m) is $\sim2$ arcmin, which is not fine enough to separate the central region from the disk.
Initially, we have checked the correlation of $L_{\rm{IR}}$ from IRAS and 24$\mu$m luminosity, $L_{24}$,  and confirmed it.
Thus we used $L_{24}$ instead of $L_{\rm{IR}}$ this time.

We defined the central and disk region of galaxies as $R_{\rm center}\leq R_{K20}/8$ and $R_{K20}/8<R_{\rm disk} \leq R_{K20}/2$, respectively.
$R_{K20}$ is the $K_s$-band 20 mag/arcsec$^2$ elliptical radius in arcsec from 2MASS Large Galaxy Atlas \citep{jar03}.
Though the 24$\mu$m disks of most local galaxies are extended to $\sim R_{K20}$, we adopted $R_{K20}/2$ as a disk extent since some galaxies were not observed in the outer regions of disk in CO.
We estimated $L_{\rm{24}}/L_{\rm{CO}}'$ of the central region and disk using AIPS task {\tt IRING} by specifying central position, position angle, axis ratio and size of disk and width of the ring to be averaged over.
The galactic global value of $L_{\rm{24}}/L_{\rm{CO}}'$ was estimated by calculating $L_{\rm{24}}$ and $L_{\rm{CO}}$ inside the $R_{K20}$.
In case that the size of CO map is smaller than $R_{K20}$, the global $L_{\rm{CO}}'$ was calculated inside the CO map.
Definition of each value is as follows:
\begin{eqnarray}
(L_{\rm{24}}/L_{\rm{CO}}')_{\rm center}&=&\frac{L_{\rm{24}}(R_{K20}/8)}{L_{\rm{CO}}'(R_{K20}/8)},\\
(L_{\rm{24}}/L_{\rm{CO}}')_{\rm disk}&=&\frac{L_{\rm{24}}(R_{K20}/2)-L_{\rm{24}}(R_{K20}/8)}{L_{\rm{CO}}'(R_{K20}/2)-L_{\rm{CO}}'(R_{K20}/8)},\\
(L_{\rm{24}}/L_{\rm{CO}}')_{\rm global}&=&\frac{L_{\rm{24}}(R_{K20})}{L_{\rm{CO}}'(R_{K20})},
\end{eqnarray}
where $L_{\rm{24}}(R)$ and $L_{\rm{CO}}(R)$ are 24$\mu$m and CO luminosity within radius $R$, respectively.

\begin{figure*}
   \begin{center}
      \FigureFile(150mm,85mm){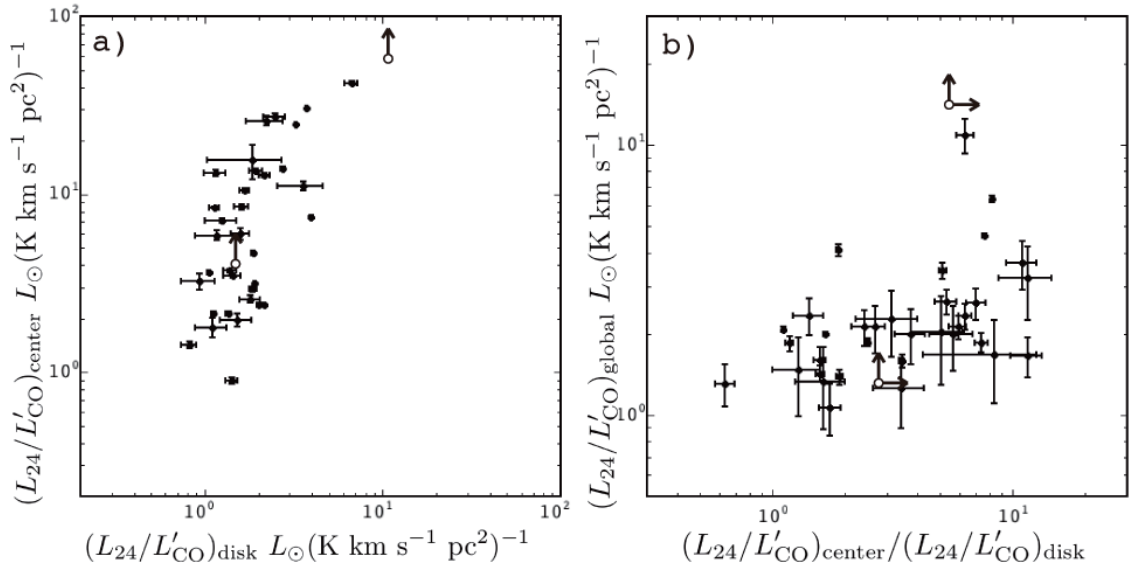}
   \end{center}
   \caption{(a) $(L_{\rm{IR}}/L_{\rm{CO}}')_{\rm disk}$ vs $(L_{\rm{IR}}/L_{\rm{CO}}')_{\rm center}$ of Nobeyama CO Atlas galaxies \citep{kun07}. Open-circle: galaxies whose central values in 24$\mu$m images are partly saturated. (b) $(L_{\rm{IR}}/L_{\rm{CO}}')_{\rm center}/(L_{\rm{IR}}/L_{\rm{CO}}')_{\rm disk}$ vs $(L_{\rm{IR}}/L_{\rm{CO}}')_{\rm global}$. Symbols are the same as (a).}
   \label{fig-5}
\end{figure*}

We plot $(L_{\rm{24}}/L_{\rm{CO}}')_{\rm disk}$ versus $(L_{\rm{24}}/L_{\rm{CO}}')_{\rm center}$ in figure \ref{fig-5} (a).
In this plot, we can see that $(L_{\rm{24}}/L_{\rm{CO}}')_{\rm disk}$ of Nobeyama samples are mainly distributed in a relatively narrow range while $(L_{\rm{24}}/L_{\rm{CO}}')_{\rm center}$ extends over almost 2 orders of magnitude.
$L_{\rm{24}}/L_{\rm{CO}}'$ value in the central region of the galaxies have more variety compared to that in the disk, which possibly reflects the difference of central activity.
From this, if galaxies have a higher value of $(L_{\rm{24}}/L_{\rm{CO}}')_{\rm center}/(L_{\rm{24}}/L_{\rm{CO}}')_{\rm disk}$, we can interpret this as because $(L_{\rm{24}}/L_{\rm{CO}}')_{\rm center}$ is high, not because $(L_{\rm{24}}/L_{\rm{CO}}')_{\rm disk}$ is low. 
In figure \ref{fig-5} (b), we plot $(L_{\rm{24}}/L_{\rm{CO}}')_{\rm center}/(L_{\rm{24}}/L_{\rm{CO}}')_{\rm disk}$ versus $(L_{\rm{24}}/L_{\rm{CO}}')_{\rm global}$.
We can see that galaxies with higher value of $(L_{\rm{24}}/L_{\rm{CO}}')_{\rm global}$ have higher $(L_{\rm{24}}/L_{\rm{CO}}')_{\rm center}/(L_{\rm{24}}/L_{\rm{CO}}')_{\rm disk}$ ratio.
This indicates that the difference in contribution from $(L_{\rm{24}}/L_{\rm{CO}}')_{\rm center}$ relative to $(L_{\rm{24}}/L_{\rm{CO}}')_{\rm disk}$ may affect the galactic global value of $L_{\rm{24}}/L_{\rm{CO}}'$.

These results imply that the variety of the galactic global value of $L_{\rm{IR}}/L_{\rm{CO}}'$ of our sample galaxies seen in figure \ref{fig-4} (b) may reflect the difference of central activity (normal, starburst or AGN) and also the difference in the degree of contribution from central star formation to the whole galaxy.
Therefore, CO observation towards galaxies at $z\sim0.1$ at high-resolution is important to understand global star formation properties.

%%%%%%%%%%%%%%%%%%%%%%%%%%%%%%%%%%%%%%%%%%%%%%%%
%%%%%%%%%%%%%%%%%%%%%%%%%%%%%%%%%%%%%%%%%%%%%%%%
\section{Summary}

We observed nine barred spiral galaxies at $z=0.08-0.25$ using the 45-m telescope at NRO, and detected CO emission with S/N $>$ 5 from six out of nine galaxies.
These are currently the IR dimmest galaxies ever detected in CO at this redshift.
The observed CO line luminosity, $L_{\rm{CO}}'$, ranges from $1.09\times 10^9$ to $1.08\times 10^{10}\ \rm{K\ km\ s^{-1}\ pc^2}$, which are similar to, or several times higher than, the average value for local spiral galaxies.
They follow the $L_{\rm{CO}}' - L_{\rm{IR}}$ relation in multiple galaxy types including local spiral galaxies, LIRGs, ULIRGs and SMGs.
$L_{\rm{CO}}'$ and $L_{\rm{IR}}$ of our sample galaxies are both higher than local spiral galaxies, though their $L_{\rm{IR}}/L_{\rm{CO}}'$ is similar or slightly higher.
These results indicate that observed barred spiral galaxies at $z\sim0.1$ form stars more actively, not because their SFE is different but because their content of molecular gas as a fuel for star formation is more abundant compared with local spiral galaxies.

This is the very first observation specialized for barred spiral galaxies in this redshift range.
However, it is hard to describe the evolution of barred spiral galaxies with only this data of samples at $z\sim0.1$, thus it is necessary to enlarge not only the number of samples at this redshift but also the samples in higher redshift.
The six CO-detected barred spiral galaxies at $z\sim0.1$ in this observation seem to be bright enough in CO to observe their distributions and have relatively similar SFE to local spiral galaxies.
These galaxies could be appropriate candidates for upcoming ALMA observations to compare the distribution of molecular gas in barred spiral galaxies with comparable resolution at different redshifts and to investigate the origin of diversity of molecular gas distributions in bars.

We would like to thank the anonymous referee for very productive comments.
We also thank the all members of NRO for observational supports.
K.M. thanks the all members of her research department in Hokkaido University, and especially to Elizabeth J. Tasker.

%%%
% See the manual for the detail.
%%%

\end{document}